# Chota Recharge and the Chota Internet
Studying the Indian mobile Internet networks

Siddharth Rao and Kiran Kumar

**Abstract**
Uniform and affordable Internet is emerging as one of the fundamental civil rights in developing countries. However in India, the connectivity is far from uniform across the regions, where the disparity is evident in the infrastructure, the cost of access and telecommunication services to provide Internet facilities among different economic classes. In spite of having a large mobile user base, the mobile Internet are still remarkably slower in some of the developing countries. Especially in India, it falls below 50% even in comparison with the performance of its developing counterparts!

This essay presents a study of connectivity and performance trends based on an exploratory analysis of mobile Internet measurement data from India. In order to assess the state of mobile networks and its readiness in adopting the different mobile standards (2G, 3G, and 4G) for commercial use, we discuss the spread, penetration, interoperability and the congestion trends. Based on our analysis, we argue that the network operators have taken negligible measures to scale the mobile Internet. Affordable Internet is definitely for everyone. But, the affordability of the Internet in terms of cost does not necessarily imply the rightful access to Internet services.

Chota recharge is possibly leading us to chota (shrunken) Internet!

# 1. Motivation: <u>Why</u> is it interesting to study the Indian mobile Internet network?

According to McKinsey & company's report [1], India has the third largest Internet user base in the world, leaving behind just China and USA. The Internet is constantly influencing the economic development, increase in GDP and overall progress of the country. With more than half of the population residing in rural and sub-urban regions, the digital inclusion achieved so far itself is a great achievement. Unlike the well-off crowd from the cities who mostly use desktop computers and fixed broadband connections to access the Internet, the rural India relies on smartphones or Internet-enabled feature phones over cellular connectivity (coverage). For the majority of the crowd (especially from the rural area) access to the Internet is only through mobile phones. Factors such as almost-ubiquitous network coverage; affordability of the devices; simplified usability with voice commands and single-touch navigations, and popularity of mobile OS/apps are helping to break the barriers of digital literacy. Most of the developing countries including India are moving towards delivering *"Internet access to Everyone, Everywhere"* through mobile networks - making it an interesting field of research. However, the rightful access to mobile Internet in India is far achieving that goal. The digital inclusion is constrained by several factors of our Internet ecosystem.

The diverse economic background along with the digital divide of urban and rural areas, contributes to the complexity of mobile Internet ecosystem of a heavily populated country like India. While the former determines the quality of mobile Internet services one can avail (e.g. smartphones that support 4G, faster data plans), the latter affects the planning and deployment of supporting infrastructure throughout the country. Due to these factors, Indian mobile Internet networks are still far behind developed countries' level [2]. On the other hand, diversity of languages and culture over the lifestyle makes it difficult to achieve unified business models (such as – using same content throughout the nation) that can benefit the Mobile Network Operators (MNOs), as well as the mobile Internet users is not as straightforward as in other densely populated countries like China. The aforementioned non-uniformities in the ecosystem lead us to categorize the hazards of achieving rightful mobile Internet access in India as follows:

## 1.1 Co-existence of multiple radio technologies

Irrespective of whether it is an urban or rural region, we get to experience different generations of radio technologies such as 2G, 3G or 4G in the same locations. Due to the humongous population and large landscape of India, rolling out a new technology by upgrading the supporting infrastructure and updating the cellular services throughout the country is a tedious and time-consuming task. This non-uniform co-existence of radio technologies is indeed a hurdle for an emerging economy like India to perceive state-of-the-art mobile Internet services.

According to ITU's country level data, 2G mobile networks are available to 96% of the world population. India is no different in this global norm with 2G being the most dominating network, 3G with 11-15% and 4G to be relatively very less connectivity [3]. Even though 2G networks are good at providing basic connectivity, they cannot support most of today's mobile Internet services. Some of the studies on reveals that the Internet usage trends are shifting towards accessing videos and media-rich websites from the basic cellular services which 2G can support. This emerging trend can only be fulfilled with 3G and 4G networks. In that realm, 2G networks are insufficient for most of our present and future Internet needs [4].

## 1.2 Lack of supporting infrastructure

Rural areas tend to have relatively sparse population (mostly with lower incomes) which is least literate about the usefulness of Internet. The cultural and language barriers lessen the demand for services Internet services in those areas. Expenses that incur due to electric power, connecting the mobile sites from rural areas to rest of the network (backhaul), set-up (up-front) and maintenance (running) of mobile sites in rural areas are three times higher than that of the urban areas. In this context, MNOs are hesitant to invest to extend their networks to fill the coverage gaps or to upgrade the existing network to support newer radio technologies in those areas. However, this impacts most of the Indian population as majority of the people are from rural areas unlike any other countries.

On the other hand, due to ever increasing movement of people towards the urban areas, the existing mobile network infrastructure is falling short to cater the demand. It seems like the operators take the urban mobile Internet users for granted as they always have better 2G coverage and fixed line connections. Hence, they do constantly upgrade their infrastructure which increases the congestion in the densely populated areas and results in poor Internet services.

## 1.3. Serving one plan to fit all

The economic status, mindset to adapt new digital skills, literacy rates and the way of living is completely different in rural parts of India than in the urban region. Due to these differences the need for Internet in rural and urban areas are not the same. As stated before, for most of the rural population, mobile phones are the first and only medium to access Internet. One can argue that the affordable mobile handset market caters the needs of having at least one low-end smartphone, the recurring cost of having an Internet pack or data plan is still not affordable for many of the low income families.

There are several models of mobile data services in developing countries – full cost data bundle, service-specific data bundle, earned data and zero-rated data [5]. In spite of lack of digital literacy, India has been one of the first countries in the world to oppose some of these models which does not support net neutrality. A recent study on exploring user experiences and perceived benefits of the aforementioned data service models [6] revealed that the people of rural India prefer full-cost models with no restrictions on websites/apps at an affordable price. We highlight again that the mobile network coverage is not optimal in rural areas and still, they are the only source for them to access the Internet. The coverage may heavily influence the speed of the perceived Internet services. They often use Internet to browse educational materials and media-rich contents.

On the other hand, most of the people from urban areas have access to Internet over *Desktop+fixed-line* broadband along with *smartphone+ mobile Internet* connections. Because of multiple options of accessing the Internet, basic data plans will be sufficient for them. They choose to use mobile medium for recreational purposes such as chat applications, and switch to the desktop for content-heavy Internet usage. So, ideally their usage pattern and need for the access to mobile Internet is totally different from the rural crowd. However, the data plans offered by major telecommunication operators are same in both the regions. At least, we could not find ourselves a tailor-made data plan separately for rural and urban regions!

### 1.4 Quality of Service (QoS) vs Quality of Experience (QoE) debate

During the initial introduction of 3G services, many MNOs speculated the sheer need of advertising the "possibility of watching buffer-free Youtube videos". Those catchy phrases and the claims still remained the same even when 4G was getting commercialized in India after 6 years. As per the recent advertisement of one of the leading MNO, they can offer 3G speed of up to 21 Mbps. This means a 10-minute YouTube video should take less than a minute to download [7]. But, we are sure that none of us have ever been able to achieve this experience irrespective of whether we live in sparsely populated hi-tech village or densely populated IT capital of India! The reasons why we cannot perceive the promised offers by the MNOs are fairly simple and straightforward.

- Yes, India is a late adapter of 3G/4G networks! So, Apart from 2g, there is no "FULL" coverage of 3G or 4G everywhere. Which means, in spite of paying for 3G or 4G data plans, there is high chances that your connections drop to 2G.
- To keep up the promise of a faster 4G connections, the MNOs are definitely working on assuring the Quality of Services implementing the supporting infrastructure here and there. However, the Quality of Experience of the mobile Internet use still remains poor and the users have not been able to perceive a good quality Internet for the extra money that they have paid for their faster data plans.
- On a first look, a high number of 4G plans were sold without the supporting infrastructure. The problem is that the initial customers may experience good 4G network for a shorter period of time. However, as the 4G user base increases, the performance or experience starts degrading due to congestion and less coverage. An optimal balance between support for a new technology and the increasing user base. The number of 4G SIM cards sold in a shorter period of time didn't seem like the planning was made efficiently.

Undoubtedly, good quality of service relies on good end-to-end infrastructure and as more users join the network, the quality is affected by the capacity constrains of the network. The QoE cannot just be measured in terms of the Internet speed. Since it is very much specific to the application being used and hence, the factors such as drop rate and delay impacts the experience of the users [4]. So, along side upgrading the network to support a new cellular technology, the MNOs should also try to optimize the user experience by adapting state-of-the art methods. Unfortunately in India, most of us do not feel a smooth mobile Internet usage!

### 1.5 'Affordability' as the sole tool for user acquisition

More than the quality of the commodity, catchy phrases like 'free', 'Cheap', 'buy one, get one', 'festival bonus' 'chota' etc. matters most to Indian consumers. Just like any other business sectors, the telecommunication marketeers have exploited the potential of selling 'affordability' to acquire more users in a shorter period of time. However, the number of 4G SIM cards sold in the previous year and the actual infrastructure that can actually support the 4G mobile networks seems like the MNOs lacked efficient planning.

Considering all the above factors, we conceptualize "**the rightful mobile Internet access**" to answer the following questions from the point of view of a regular Indian Internet user:
1. If I have paid for 4G (or 3G) data plans, am I getting the services what I have paid for? Simply speaking, am I getting the optimal download speed which I should be getting?
2. If I have paid for 4G (or 3G, may be 2G as well) data plans, are my services and experience of using those services hindered by coverage, congestion, infrastructure and interoperability between the MNOs.
3. What exactly am I promised by the MNOs when I pay for a specific data plan?
4. Is there anything that I could do from my side (e.g. upgrading my mobile handset, switch to better data plans) experience better mobile Internet services?
5. How is my MNO doing in comparison with other MNOs? Will switching to another MNO improve the quality of my mobile Internet experience?
6. Has my MNO taken enough measures over the past years to improvise their services?

Unless the answers to the above mentioned questions are positive, we argue that the chota recharge that Internet users do is possibly leading us to a chota (shrunken) Internet. Hence, they possibly are unaware of the fact that they are not getting what they have paid for in order to rightful access and use the mobile Internet.

As far as we know there are no 'complete' and 'open' datasets available to study the performance of MNOs. Especially for a country like India which is booming up with mobile Internet, such a study is much needed to define policies and strategies to defend the digital rights before it is too late. In this realm, we discuss the methods to follow and the factors to should be considered while studying the mobile Internet measurements. We provide the usefulness of such a study by demonstrating some of the results (or say 'polite claims') that we attained using the same research methodology from an academic (close) dataset that we had access to. We strongly believe that, these initial findings may open up the quest to curate openness, transparency and public knowledge in the mobile Internet ecosystem.

## 2. Methodology: <u>How</u> to study the mobile Internet ?

There are various scientific articles emphasizing the research on user-perceived mobile Internet measurement data such as Paul *et.al.* [13] in which analysis of the network traffic has been done from base station's point of view (i.e. with the cooperation of the MNOs). Works by Tan *et.al.* [14] shows how the capacity and performance of 3G networks has been analyzed on a global scale. In India, it might be difficult to work on an open data project which publicly showcases the performance and issues of MNOs with their cooperation. This is because the MNOs consider it to be a business and strategic secret which is not be disclosed publicly. However, from public interest's point of view to know the performance of the MNOs, it is suitable to perform crowd-sourced measurement data collection and analyze it further. The advantage of crowd-sourcing (such as one of the works done to analyze the cellular connectivity in India [15]) is that it can be done without the cooperation of the MNOs. However the success of such experiments completely rely on the participation of more number of people and the quality of data collected.

## 2.1 The crowd-sourced generic datasets

Irrespective of the cooperation of the MNOs, the analysis boils down to the exploration of TCP traffic parameters such as download/upload throughput (speed), latency, congestion, etc. Instead of designing an app which blindly collects and displays the download/upload speeds like any other speed test apps, a sophisticated app for researching the mobile Internet should collect the following parameters:

- Location of the mobile from GPS network
- Latency
- Manufacturer, model, operating system and version
- Network and subscriber operator
- Signal strength
- Location of the base station
- Mobile technology (such as UMTS, HSPA)
- IP address and transport ports
- Timestamp

Due to its simplicity, download speed (more technically it is downlink throughput) is globally accepted as an indicator of Internet access. Even we use the the same to explain it to the general public. However, the download speed is affected by factors such as mobile device (hardware), radio technology, subscription/data plans, network coverage, and the congestion caused by other users [10]. The thumb rule of making sense of mobile Internet measurement data is to analyze "*which of those factors have particularly limited the download speed*" (Or in other words, which factor hindered the better or rightful mobile Internet access). Mobile device, radio technology and data plans are considered as *artificial limiting factors* as they can be determined even before taking any measurements. Whereas coverage and the congestion have a tendency to occur randomly and they constantly vary depending on location and time of the day. Hence, they are considered as the *natural limiting factors*.

**Device**: The network speed achievable by a specific mobile device is limited by factors such as the operating system, implementation of the network stack, design of the antenna and the hardware (or chip-set). The maximum download speeds of the best (high-end) device is as high as 5 times as that of the worst devices in comparison. For each mobile device, its manufacturer provides the maximum achievable throughput per technology (which can be considered as the theoretical upper bound). Irrespective of how ideal the network condition is (such as low congestion, good signal strength, better data plans, support for a specific radio technology by the connected base station), a specific device cannot achieve a throughput beyond its theoretical upper bound.

**Radio technology**: Different generations of radio technologies such as 2G, 3G and 4G 1 are the standards for wireless communication of high-speed data for mobile phones. Each of them are based on improvements from their predecessors in terms of capacity, speed using different radio interfaces and core network. Such radio technology standards are developed by 3GPP, which also specifies the theoretically maximum achievable throughput (upper bound) for each of those standards per release.

**Data plans**: 'Data plans' are simply the download speed or the data limit (in developing countries) promised by the MNOs for certain amount of money. In developed countries (e.g. Finland, the mobile operators specify the maximum throughput that a user can perceive by paying a fixed amount of money. In these cases, the amount of mobile Internet data that the user is allowed to use is often unlimited. On the other hand, in the case of India, the mobile operators specify the the maximum amount of Internet data that a user can use in a fixed period of time by paying a fixed amount of money (e.g 100 Rs of recharge for 1GB of 3G Internet data). We consider the maximum speed promised per data plan as the theoretical upper bound.

For any measurement, if the throughput falls within a window between the upper bound and a threshold value of the aforementioned artificial limiting factors, we conclude that the factor in consideration is the limiting factor for the low speed Internet. Apparently, there is nothing much that can be done from MNO's side as these are the limitations due to existing technologies or hardware. So, the natural limiting factors can be ignored due to its dependence on newer scientific methods rather than mere tweaking or investments from the MNOs.

**Congestion**: The presence of congestion in any measurement can be identified by observing more fluctuations in the download samples. The unnatural spikes in those fluctuations account for packeting reordering in TCP connection of the samples. Since, these packet reordering will give biased results about the throughput when we correlate the throughput in terms of congestion, we normalize the affect of packet reordering occurrences by filtering out each of those samples by considering the moving average. We then calculate the mean and Relative Average Deviation (RAD) of each window for these filtered samples to identify the upper bound for congestion, which is nothing but the window with highest mean value and lowest RAD. Furthermore, we calculate Mean Absolute Percentage Error (MAPE) for each window (except for the windows pertaining to TCP slow start) by considering the aforementioned upper bound as expected value and the samples of the window as actual values. Based on our exploratory analysis we have determined threshold ranges for classifying the congestion into low, medium and high pools based on MAPE. These threshold ranges can be fixed in accordance with the need for accuracy while classifying the congestion. However, for our ease of analysis, we have considered MAPE < 10% as 'low', 11% - 25% as 'medium' and beyond 26% as 'high' congestion.

*Note 1: We consider RAD as it is directly proportional to the fluctuations in a window. Lower RAD signifies stable windows.*

**Coverage**: It is not possible to find upper bound on coverage by any similar methods as described before. Some of the researchers have tried to deduce such bounds using signal strength. However, signal strength is not an accurate indicator for correlating coverage with throughput [11] and hence, it is not advisable to consider any upper bounds directly for coverage. Instead, the best approach is to correlate coverage in terms handovers which simply refers to change of base station.. It is the scenario where a mobile device goes beyond the coverage (and hence the proximity of a base station), thereby entering the coverage of another base station.

In practice, throughput is heavily influenced by the above mentioned natural limiting factors. So, it is advised to focus mainly on them to deduce analogy about network performance. The modus operandi for considering specific data samples from the historical dataset containing the TCP traffic parameters that we considered is as follows:
- Ignore those measurements which are limited by the artificial factors.
- Correlate the network performance in terms of natural limiting factors which are grouped by radio technology where and when applicable.

**2.2 Scenario-based datasets**

Though adequate measurement data can be sought from a generic dataset as before, to get a better understanding of the user-perceived network quality and conduct an in-depth analysis of performance of mobile networks at different times of the day, we need to improvise the quality of the data by adding some scenario-based data points. This is because, from user's point of view, if and only if the users think that they are experiencing poor Internet services, they use their speed test apps to confirm the same. Due to this mentality, the data points in a generic dataset may not be technically random enough and also, it might miss some important scenario. These datasets can be collected either from controlled subjective experiments done using measurement campaigns or from an automated machine which takes continuous measurements. Two of the must and should scenarios are:
- *Continuous measurements for a single base station around the clock*: This experiment enriches the data in terms of hourly breakdown of performance and variation in congestion of particular base stations. Eventually it allow us to capture the performance drops and technology degradation due to high congestion (during busy hours of the day).
- *Continuous measurements for several base stations connected during commute*: Through this experiment we can capture the overall network deployment pattern in specific vicinities. Moreover, the data required to analyze the performance of the network (in terms of performance drops and technology degradation) during each base station change along with the possible correlation to handover and roaming can also be obtained.

<u>Note:</u> *Continuous measurement refers to measurements taken continuously at fixed intervals of time.*

Once the dataset is rich enough to evaluate, we can analyze overall network performance by considering download speed as a Key Performance Indicator (KPI) for reasoning the network deficiencies, as it is one of the most widely used metric for evaluating QoS. However, all the data that is gathered in the dataset as per 2.1 and 2.2 is indeed measured by the end-users, which potentially gives a better understanding of QoE. While considering user-perceived measurements, download speed alone cannot be an accurate KPI, as the variations of throughput heavily influences the perceived quality of end-user. Hence, we should emphasize on congestion and coverage mainly to correlate with the speed, as it takes user-perceived factors into consideration while providing a comprehensive overview of network performance in terms of QoS.

## 3. Results: What are the usefulness of studying mobile Internet measurements?

Now, we discuss some of the early results that we found by analyzing a dataset that we had access to. We used a mobile Internet measurement app called Netradar and used the data collected and curated by it for our analysis. Netradar system [8] was launched in June 2012 in order to acquire scientifically reliable data about mobile networks and devices. Contrary to other similar services which simply report network speed and latency, Netaradar collects a number of contextual parameters to perform series of network related measurements and analysis to understand the performance better. The app collects the following parameters for every measurement: location from GPS or network; download and upload speeds; latency; manufacturer, model, operating system and version of the device; network and subscriber operator; signal strength; base station; mobile technology, such as UMTS, HSPA; IP address and transport ports; timestamps. To date, the data collected by Netradar accounts for measurements from 175 countries covering 946 different operators, 8216 unique mobile devices, 278142 unique users have installed the app and have contributed to the crowd-sourced measurements. There were 8807 unique users from India which accounts for approximately 40000 measurements. The dataset also includes the user experiment campaign [9] that we launched in order to fill the missing gaps of Internet usage. Unfortunately the dataset as well as the Netradar application is not open. However, due to its solid academic grounds [10], we assure the correctness of the data curated by the app.

We understand that the data that we used for analysis may not depict the exact situation of the mobile Internet ecosystem of India. We also understand that it may or may not accurately provide a generic big picture of the performance metrics of the mobile Internet. However, we claim some of the deficiencies that we could point out just based on the dataset that we had. This is just to unleash the usefulness of procuring such a dataset and analyzing it further to explore more. We appreciate the efforts by leading operators such as Airtel for their Open Network initiative [11] to be transparent. Indeed we would be more than happy to see such initiatives by other operators as well in future. However, in terms of academic research we wanted more parameters to analyze and hence, we used the the Netradar dataset in context to come up with research methodology guidelines provided in this essay. Based on our analysis, we now put forth some of our 'soft claims' and justify it. However, more in-depth investigation of the following claims is an open research question and a collaboration opportunity between academicians, telecommunication industry and policy makers.

**Claim 1: Overall mobile data networks in India are below the acceptable standards.**
Analysis on the overall performance in terms of download throughput of mobile networks is shown in figure 1. This histogram shows the distribution of throughput from several places across India including urban, suburban and rural areas. The right-tailed plot clearly shows that the measurements which have recorded better performances are relatively very few. The majority of the measurements have recorded download speed less than 1Mbps which is fairly low compared to 3G standards. Irrespective of the MNOs and the actual site of measurement (rural or urban), a country where 3G services are around for nearly 8 years, the so-called 3G network doesn't seem to deliver performance according to the global norms.

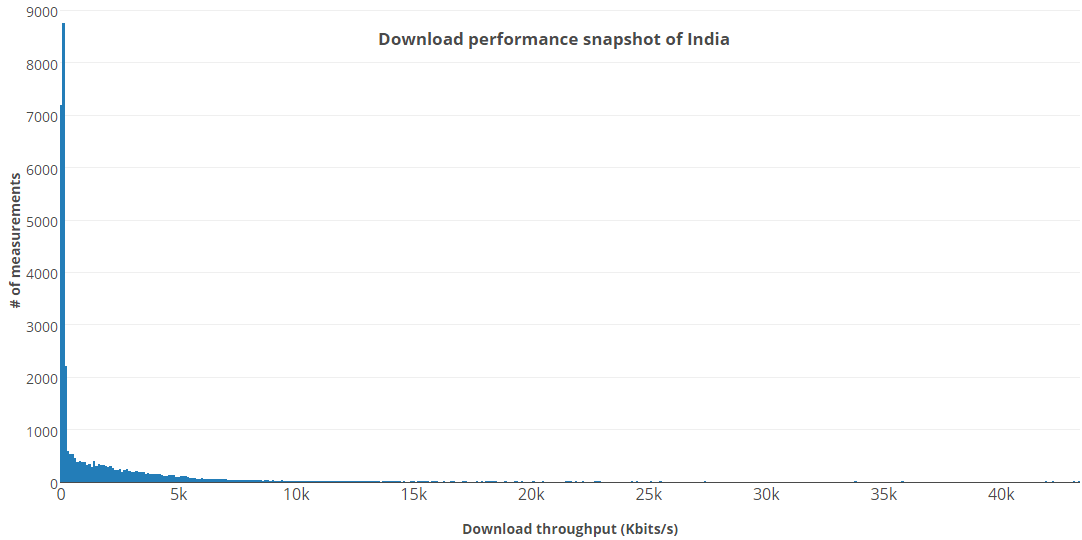

Figure 1: Overall performance of Indian mobile Internet

When we further explored the reasons for the overall poor performance of the Indian mobile networks, we figured out that it is due to the inefficient management of the base stations (cell towers). For better understanding, we plotted (refer figure 2) the average download speed of four base stations through which we had the continuous measurement data throughout the day.

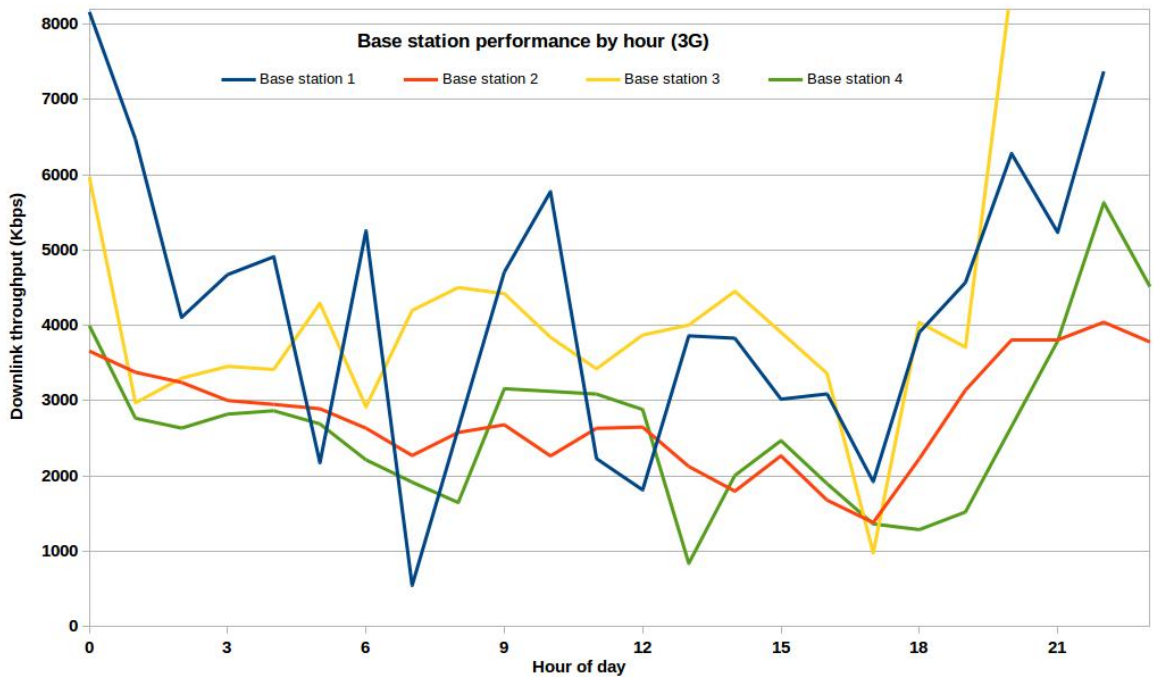

Figure 2: Performance of base stations throughout the day

The graph in context (figure 2) shows a consistent dip in speed/throughput during peak (busy)

hours of the day (e.g. from 7 AM to 5 PM). It was found that on an average the speed in urban region is decreased by 60% during the busy hours compared to that of the non-busy hours. These base stations were geographically located near one of the main bus stations, which is always crowded by daily commuters during the busy hours. Though the reasoning for this performance is due to congestion - which is fairly straightforward, 60% drop in the speed again makes questionable to accept is as a 3G network according to the definition.

**Claim 2: Mobile network operators are failing to provide the required quality of service to their users**

The figure 3 represents the evolution of 3G and 4G technologies in terms of average performance (download speed) over the past 3 years grouped in terms of each quarter. Due to uniform distribution of economy and penetration of sophisticated internetworking technologies, we considered similar measurements from Finland as a standard bench mark to perform a comparative analysis with Indian mobile networks.

In this time series graph we are more interested in the pattern of the curve for each country rather than the mere difference in the download speed value. It is already well established fact that the Indian mobile network lags behind the Finnish network, which is one of the best mobile networks of the world.

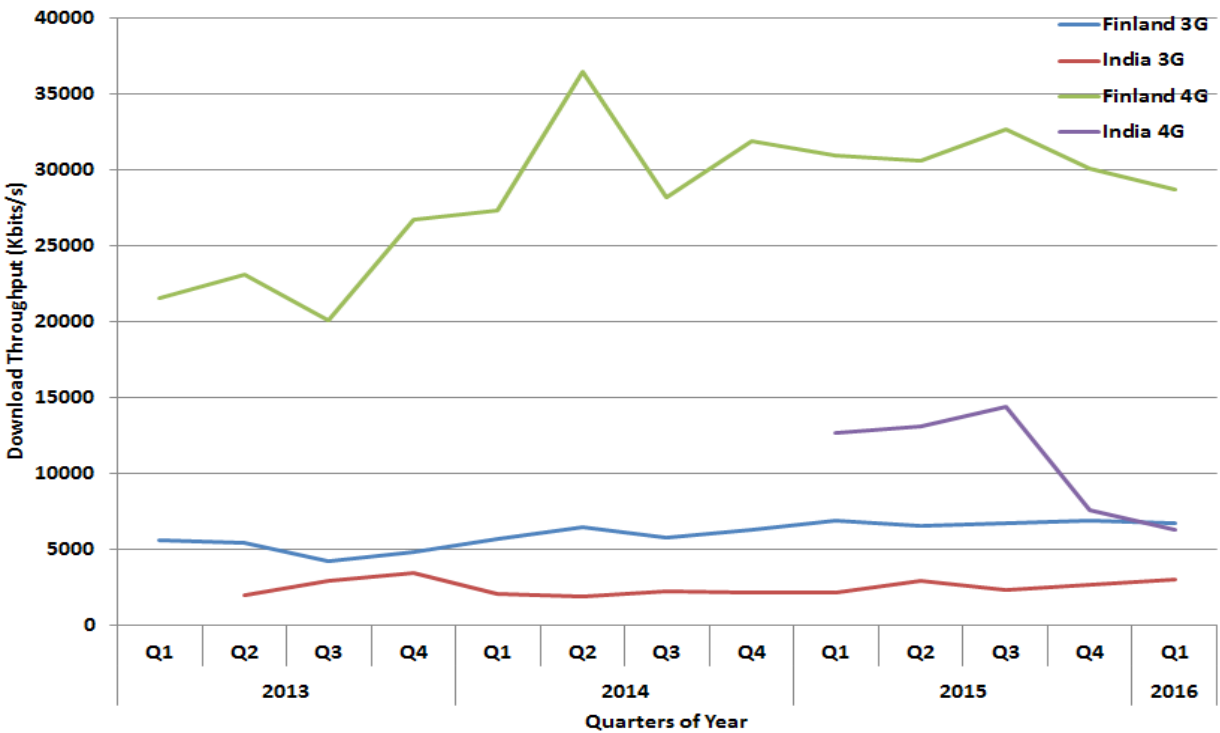

Figure 3: Comparison of performance between Indian and Finnish mobile Internet Networks.

In case of Finnish 3G networks, we can observe a drop (in Q3 2013) followed by a rise and consistent stable slope. We speculate this behavior as follows: the drop here represents the possible increase in 3G customer base, whereas the rise and stable slope is the response by the

MNOs in Finland to optimize the network services to support the increasing customer base. Even though a similar drop pattern in seen in Indian networks suggesting an increase in 3G customer base, the performance has become stagnant with negligible rise till date. This pattern could be because of two reasons:
1. The operators in India are not optimizing the network services like in Finland;
2. Even if the network operators are improvising their services, it is not sufficient enough to cope up with the increase of demand of 3G services from Indian customer base.

In case of 4G network performance, Finland being the earliest adopter of 4G commercial network services, it showcases a smooth evolution. The steep drop in case of Indian 4G networks is a possible indication of excess acquisition or upgrade of 4G customers without having enough supportive infrastructures. At this point we do not speculate anything further on Indian 4G networks, as India is one of the recent adopters of 4G services and this fact is evident by the relatively less number of 4G measurements in our dataset.

The figure 4 also shows the evolution of network performance (in terms of maximum and average performance achieved) over the past 3 years, but by taking into account the 2 major commercial network types – cellular and WLAN. The huge gap between maximum and average speeds between these network types indicate the under-performance of cellular networks. However, the high speeds achieved over WLAN is an indication that the existing back-end networks are strong enough and it could be possible that they are fine-tuned to prioritize the fixed-line connections. MNOs may need to take accountability to improvise Internet connections over cellular (mobile) networks, as that is what India at the moment need to achieve good rate of digital literacy.

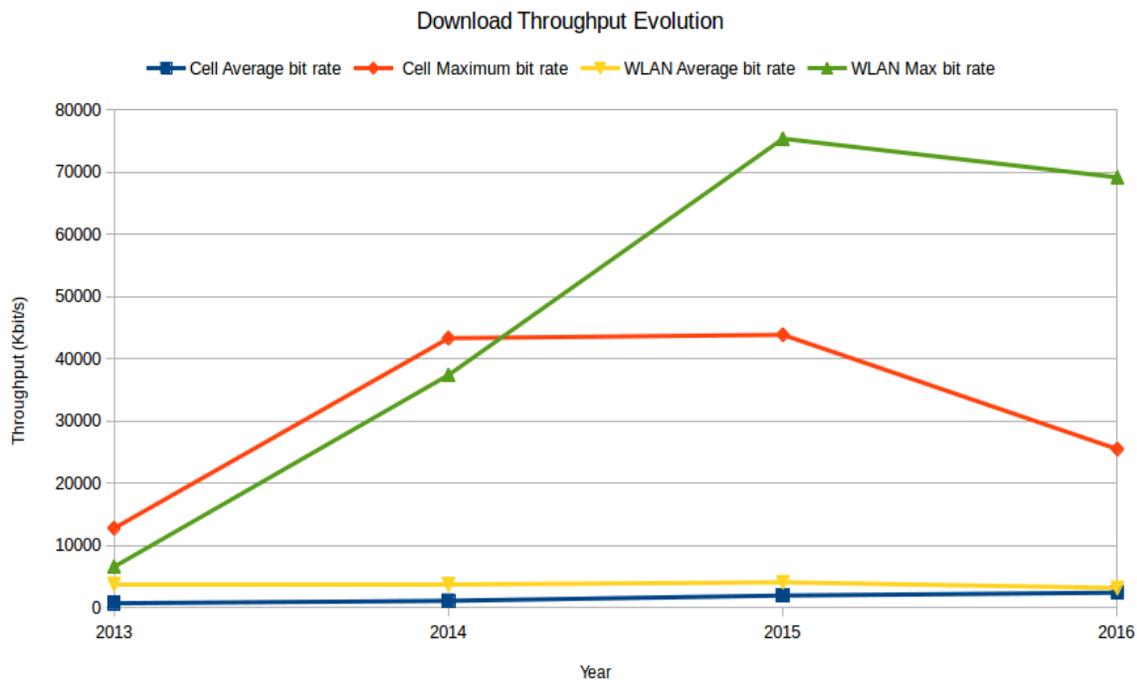

Figure 4: Comparison between cellular (mobile) Internet and fixed-line Internet speeds

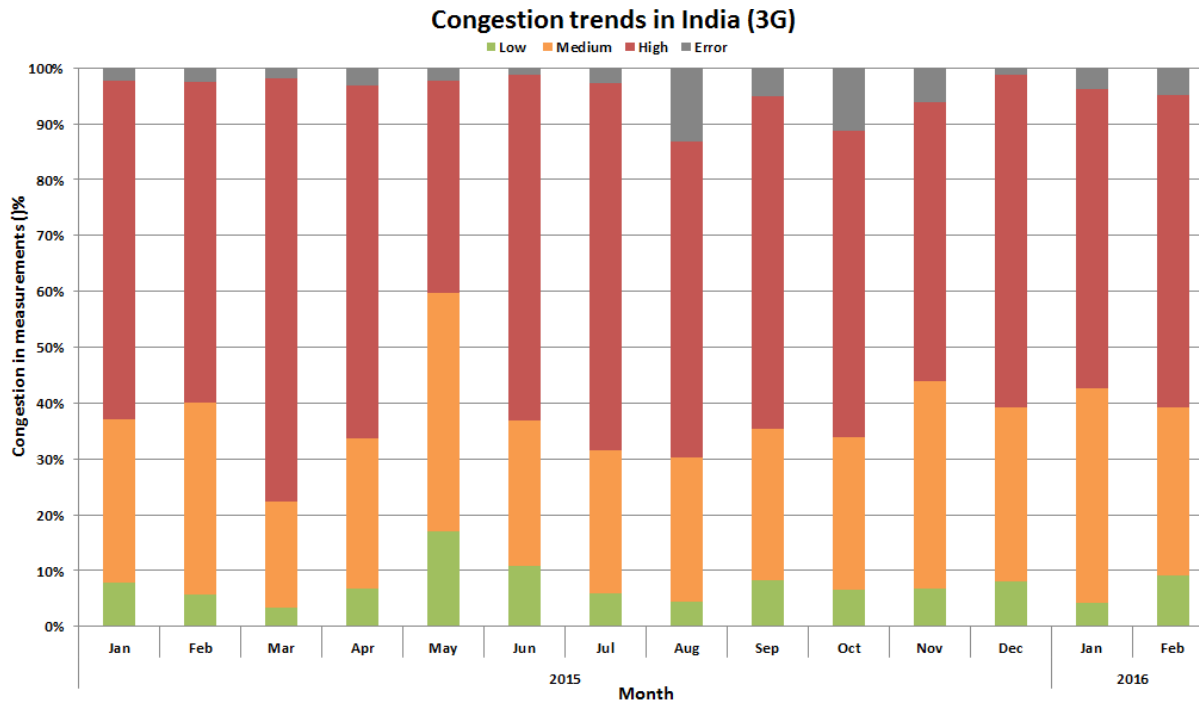

Figure 5: Congestion trends over the past years.

The congestion trends in India for 3G base stations for the year 2015 is represented in figure 5. Using this analysis, we wanted to understand the measures taken by major operators in India to address the issues pertaining to congestion (as discussed in claim 1), thereby their support for improvising throughput that an end-user can perceive. We found that, over past one year (or even before that) 90% of the measurements in the urban areas consistently fall into the categories of high or medium congestion. This trend potentially indicates that the operators have taken negligible measures to address the congestions issues over the past years.

Figure 6 shows the performance of some major Indian MNOs, where the box plot shows the spread of the download speed per operator. The distribution of the measurements for all samples is restricted far below 2Mbps and this indicates that none of the operators are up to their mark in providing 3G services. Comparatively, Airtel and Vodafone show some promising speeds at least for some measurements.

Figure 7 represents an extended version of the analysis of figure 6, where the operators' performance is averaged out by the hour of the day. There is a clear distinct between the performance of operators during the busy hours and the rest of the day. This could be obvious and common in every countries. However, irrespective of the fact that all the operators have the same dip pattern during the peak hours, the magnitude of deprecation due to the capacity and congestion aspects of the network is an interesting factor to study. Even in this graph, Airtel and Vodafone seems to show better numbers compared to its counterparts.

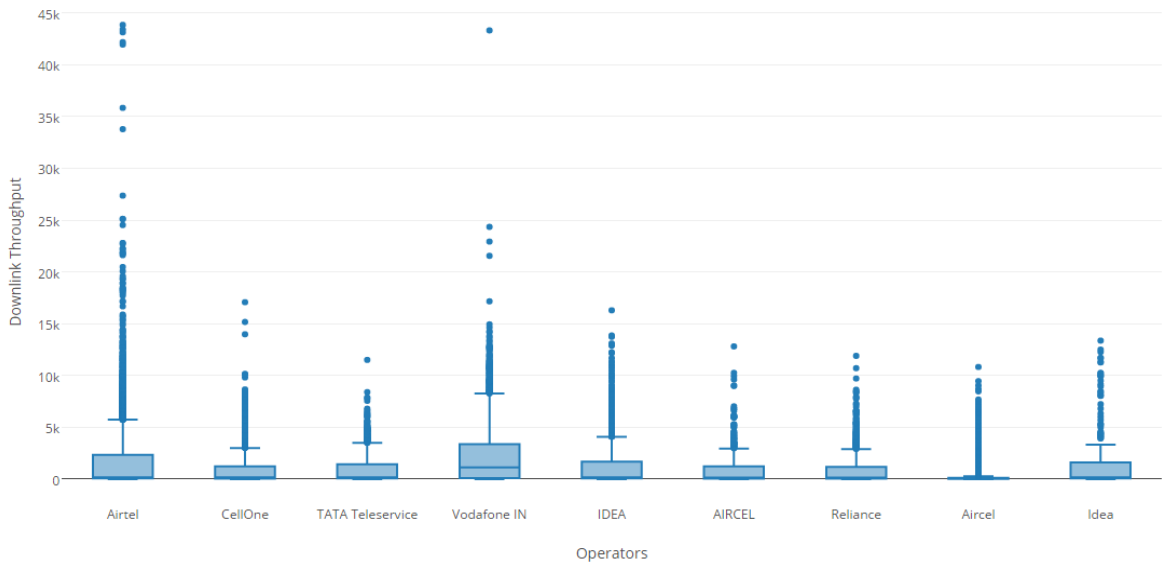

Figure 6: Performance according to different operators.

Figure 8 shows the distribution of measurement samples grouped in terms of technology. The numbers clearly fall below the range of the standards specified by IMT-2000 for 3G and IMT-Advanced for 4G. The major chunk of LTE measurements fall below 15Mbps, which is quite low even for 3G standards. The underlying technology may have 4G support, but providing the network capacity to achieve the speeds is required and there is a lot of scope for improving the mobile data services from the operators end.

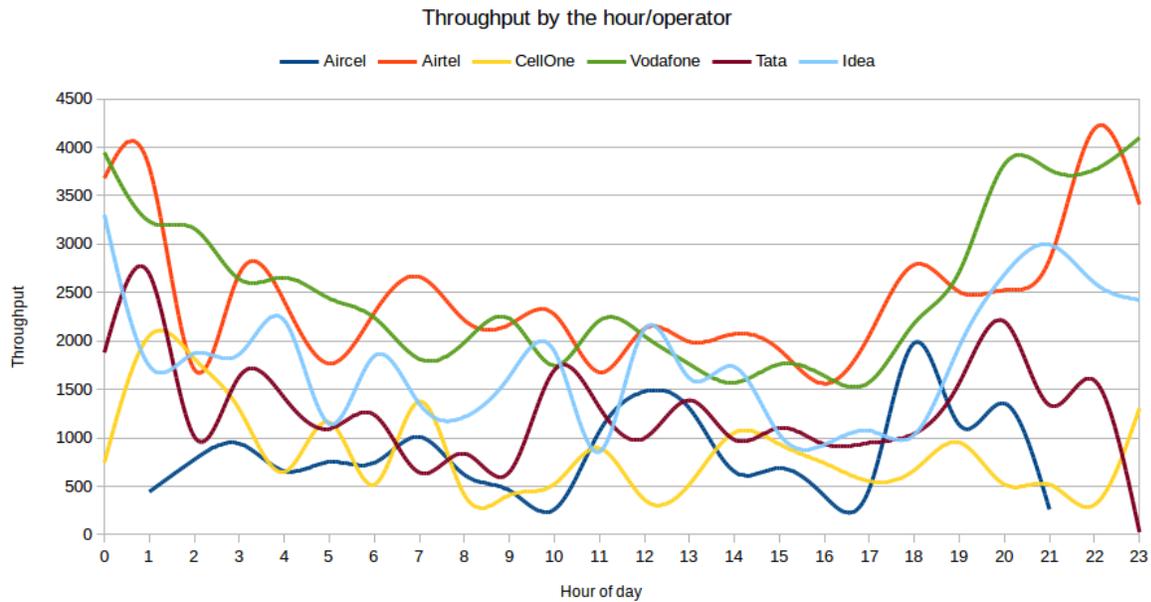

Figure 7: Performance of different operators throughout the day

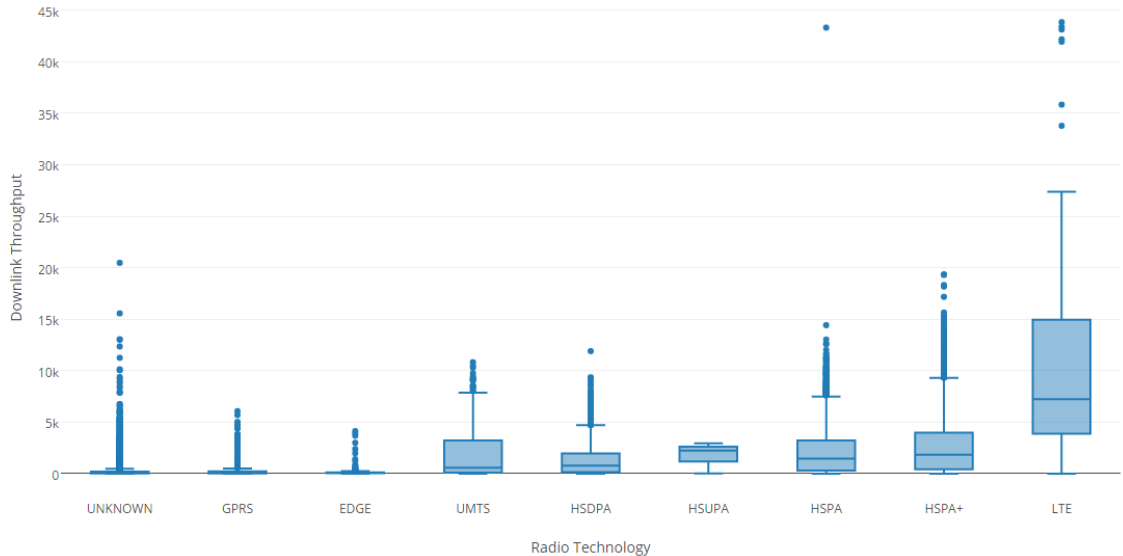

**Figure 8: Distribution of measurement samples grouped by technology.**

*Note*: Apart from the representation in figure 8, for the sake of simplicity, throughout this article we have grouped EDGE and GPRS measurements as 2G; 3G as a broader and generic term for UMTS, HSPA and HSPA+ measurements and 4G/LTE is considered as it is.

**Claim 3: Good Signal strength – A myth**

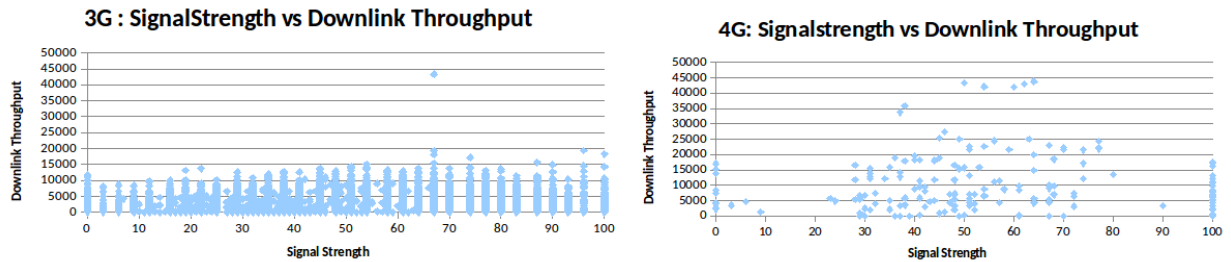

**Figure 9: Signal strength vs Download throughput in 3G/4G networks.**

Figure 9 shows the correlation between signal strength and download throughput. It can be seen that there is now pattern of increase on throughput with signal strength. This is a common misnomer among people that good signal strength yields good performance, which is not true in the case of mobile networks [11]. So having good coverage and thereby good signal strength promised by the MNOs may not help to perceive better speeds while using mobile Internet. After certain threshold, the signal strength will become obsolete in its contribution to better and rightful Internet access.

**Claim 4: Coverage holes and the interoperability issues**

Figure 10 shows the user experience of a single user over a period of 5 months in rural area. In this scenario, the user has recorded over 3000 measurements at different time of the day and at different locations (during his regular commute) over a period of 5 months and the download throughput for each measurement has been plotted in the graph. The user has used subscription from 2 different operators - Vodafone and BSNL during these measurements. From the measurements, although it is clear that the user has subscribed to 3G connection from both the operators, the 3G speeds observed over the BSNL network is lower compared to the 3G speeds of Vodafone network. It is also evident that BSNL has downgraded its technology from 3G to 2G in most of the cases. However, Vodafone seems to be consistent in its technology and speeds. An interesting observation that can be made from the graph is that, during the initial 2-3 months Vodafone has provided a consistent download throughput in the range of 3.5Mbps to 4Mbps. During the last 2 months, the download throughput speeds have increased to 5Mbps to 6Mbps.This indicates some level of improvement from the network operator perspective. As a whole, it is evident that BSNL's network is very poor in comparison with Vodafone's network in the locations where this particular user uses data over the mobile network. Also, it is clear that the user is more inclined to use the Vodafone subscription than BSNL due to the coverage issue.

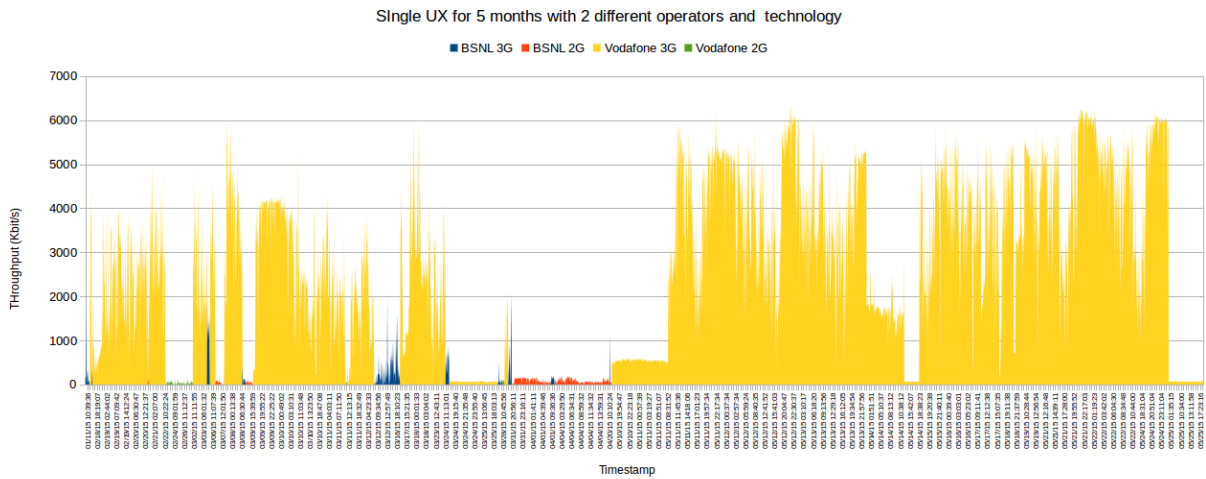

Figure 10: Continuous measurements taken during commute for 5 months

Figure 11 represents the variations in performance with user's movement for a specific operator (say operator A). We found that the download speed drops during every handover of the network from one base station to other. What is more interesting than the obvious absence of smooth handovers is the sudden drop in the speed (as seen in the center of the graph) where the radio technology was downgraded to 2G from 3G. Additionally the frequency of 2G handovers were more compared to that of 3G handovers. Based on the frequency of handovers and the supporting base stations attached during the handovers, we can speculate the patterns of 2G and 3G base station deployment in a special geographic location where the participant has conducted his measurements.

We analyzed the coverage of another operator (say operator B) in the same geographical area of the measurement data from the previously mentioned analysis of operator A's network. Apparently operator A and B had no roaming agreements with each other or with other operators providing service in the same coverage area. In this case we found that the speed was decreased by ∼90% in case of operator B when compared to operator A. In all those cases where the throughput was decreased, the signal strength was also decreased by ∼50%. By comparing the difference in performance of the operator's networks, one can infer about the supporting infrastructure and deployment patterns of the operators to speculate their interoperability while providing service in the same coverage area.

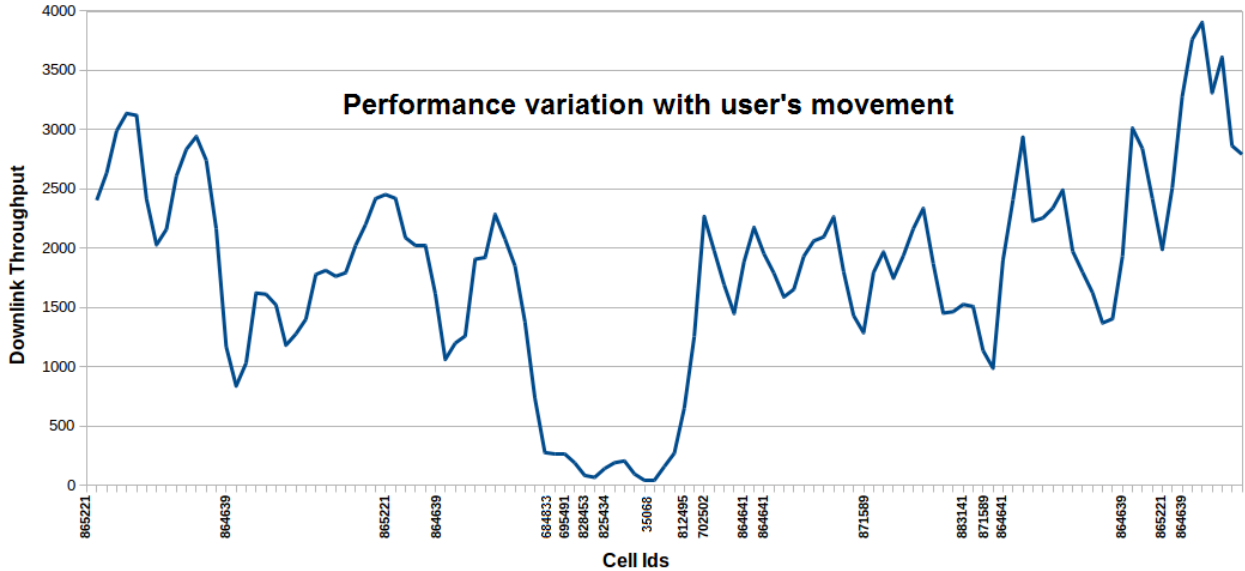

Figure 11: Continuous measurements taken during commute

These coverage holes for 3G network for operator A could have been easily resolved by having appropriate roaming agreements with Operator B. However, unlike in the developed countries sophisticated infrastructure sharing [12] and roaming agreements are not so common in India. Even if they exit, there is no public information available regarding the same.

## 4. Conclusion

Irrespective of whether it is urban or rural region, the technology penetration (existence of 2G/3G and 4G networks) can be seen throughout India. Due to high population density in urban areas especially during busy hours, mobile networks are congested and and this severely decreases the overall throughput of the country. Congestion occurred at the base stations can be used as an indicator of performance of an operator's infrastructure, meanwhile considering several other factors such as capacity (i.e the maximum number of subscribers that a base station can serve). Even though the capacity of base stations cannot be accurately predicted [14], congestion is directly correlated to capacity. Persistent issues related to congestion over a period of time indicates that the capacity provided by the operators is not sufficient enough to cater highly populated areas of India. Even if they have improvised their infrastructure to accommodate more users per base stations, their incapability to cope up increasing number of users is an early indication of lack of network planning. Based on the frequent change of 2G base stations compared to that of 3G, we can infer there are more number of 2G base stations deployed, in spite 3G networks being equally popular as 2G. We found several cases 6 of 3G coverage holes where the network was downgraded to 2G due to sparse deployment of 3G infrastructure. Additionally, our analysis revealed that nearly 50% of the time, a 4G user in India is connected to 3G or 2G base stations which is again due to sparse deployment.

To improvise the performance of operator's infrastructure, they have to either include sophisticated congestion pricing models or congestion control mechanisms as part of their future network planning. Like the developed countries, if the operators from developing countries adopt congestion pricing models such as time-of-day pricing, they might loose their market share due to economical disparity amongst their users. Other models such as pricing per application were heavily opposed by majority of the mobile users and also criticized by the constitution as those models obstruct net neutrality. On the other hand, operators from developed countries can afford expensive congestion control mechanisms like over dimensioning (i.e. deployment of more infrastructure), developing countries cannot scale such mechanisms due to large population and high capital expenditure along with inefficient usage of network capacities. A better approach would be to incorporate techniques such as cell breathing [17] to utilize the existing infrastructure, and diminish the effect of congestion or coverage on network performance. To address the coverage and interoperability based shortcomings, the operators from developing countries must include the following two measures. First of all, inter-operator roaming agreements should be improvised. As we stated earlier, roaming agreements ensure better quality of services even when there is a coverage hole for specific operators. Secondly, Mobile Virtual Network Operator (MNVO) models should be implemented. In both the cases, the operators can optimize their existing infrastructure and utilize them to provide better services to wide range of mobile Internet users. Furthermore, such infrastructure sharing models not only reduces the cost of service delivery, but also they encourage innovative telecommunication services. These two approaches have been one of the key success factors of better quality of network in developed countries [16].

In future, mass deployment of Internet of Things would completely reverse the performance metrics of Internet measurements, as these technologies heavily depend on uplink throughput in contrast with all the present day analysis emphasizing on downlink throughput. In such situations, having a robust back-end network support is highly required. However, developing countries like India demands lateral design paradigms in practical realization of sustainable IoT centric networks [18]. In this realm, we strongly believe that the analysis we presented about the evolution and adaptation of current networking technologies could potentially benefit while speculating the future trends.

**References:**
[1] McKinsey (2013) *India's Internet opportunity*. Available at: http://www.mckinsey.com/industries/high-tech/our-insights/indias-internet-opportunity (Accessed: 18 August 2016).
[2] Business Wire (2015) Indian mobile Internet networks still far behind developed countries' level. Available at: http://www.businesswire.com/news/home/20150114005139/en/Netradar-Aalto-University-Indian-Mobile-Internet-Networks (Accessed: 18 August 2016).
[3] Sarkar, D. (2016) *The Times of IndiaTech - 5 reasons why 4G is not able to excite mobile users in India*. Available at: http://timesofindia.indiatimes.com/tech/tech-news/5-reasons-why-4G-is-not-able-to-excite-mobile-users-in-India/articleshow/50918140.cms (Accessed: 18 August 2016).
[4] Wu, P., Jackman, M., Abecassis, D., Morgan, R., De villiers, H. and Clancy, D. (2015) *State of Connectivity 2015: A report on global Internet access | Facebook newsroom*. Available at:

http://newsroom.fb.com/news/2016/02/state-of-connectivity-2015-a-report-on-global-internet-access/ (Accessed: 18 August 2016).

[5] The Alliance for Affordable Internet (A4AI) (2015) *The Impacts of Emerging Mobile Data Services in Developing Countries - Models of mobile data services in developing countries*. Available at: http://1e8q3q16vyc81g8l3h3md6q5f5e.wpengine.netdna-cdn.com/wp-content/uploads/2015/11/MeasuringImpactsofMobileDataServices_ResearchBrief1.pdf (Accessed: 18 August 2016).

[6] The Alliance for Affordable Internet (A4AI) (2016) *The Impacts of Emerging Mobile Data Services in Developing Countries - Mobile Data Services: Exploring User Experiences and Perceived Benefits*. Available at: http://1e8q3q16vyc81g8l3h3md6q5f5e.wpengine.netdna-cdn.com/wp-content/uploads/2016/05/MeasuringImpactsofMobileDataServices_ResearchBrief2.pdf (Accessed: 18 August 2016).

[7] Kaushik, M. (2015) *Distress Call: What is leading to the death of 3G in India?* Available at: http://www.businesstoday.in/magazine/trends/factors-that-are-contributing-to-death-of-3g-in-india/story/220551.html (Accessed: 19 August 2016).

[8] *Netradar*. Available at: http://www.netradar.org/en/about (Accessed: 21 August 2016).

[9] Rao, S.P. and Kumar, K.M. (2016) *Netradar India: A User Study*. Available at: http://netradar-india.github.io/ (Accessed: 21 August 2016).

[10] Sonntag, S., Manner, J., & Schulte, L. (2013, May). Netradar-Measuring the wireless world. In *Modeling & Optimization in Mobile, Ad Hoc & Wireless Networks (WiOpt), 2013 11th International Symposium on* (pp. 29-34). IEEE.

[11] Sonntag, S., Schulte, L., & Manner, J. (2013, April). Mobile network measurements-It's not all about signal strength. In *2013 IEEE Wireless Communications and Networking Conference (WCNC)* (pp. 4624-4629). IEEE.

[12] Groupe Spécial Mobile (GSM) Association (2012). *Report - Mobile Infrastructure Sharing*. Available at: http://www.gsma.com/publicpolicy/wp-content/uploads/2012/09/Mobile-Infrastructure-sharing.pdf (Accessed: 19 August 2016).

[13] Paul, U., Subramanian, A. P., Buddhikot, M. M., & Das, S. R. (2011, April). Understanding traffic dynamics in cellular data networks. In *INFOCOM, 2011 Proceedings IEEE* (pp. 882-890). IEEE.

[14] Tan, W. L., Lam, F., & Lau, W. C. (2008). An empirical study on the capacity and performance of 3g networks. *IEEE Transactions on Mobile Computing*, *7*(6), 737-750.

[15] Koradia, Z., Mannava, G., Raman, A., Aggarwal, G., Ribeiro, V., Seth, A., ... & Triukose, S. (2013, December). First impressions on the state of cellular data connectivity in India. In *Proceedings of the 4th Annual Symposium on Computing for Development* (p. 3). ACM.

[16] Shin, D. H., & Bartolacci, M. (2007). A study of MVNO diffusion and market structure in the EU, US, Hong Kong, and Singapore. *Telematics and Informatics*, *24*(2), 86-100.

[17] Mishra, A. R. (2004). *Fundamentals of cellular network planning and optimisation: 2G/2.5 G/3G... evolution to 4G*. John Wiley & Sons.

[18] Misra, P., Simmhan, Y., & Warrior, J. (2014). Towards a Practical Architecture for India Centric Internet of Things. *arXiv preprint arXiv:1407.0434*.